\documentclass[twocolumn,showpacs,preprintnumbers,amsmath,amssymb,aps]{revtex4}

\usepackage{graphicx}
\usepackage{dcolumn}
\usepackage{bm}
\usepackage{subfigure}

\def\x{{\mathbf x}} 
\def\z{{\mathbf z}}
\def\y{{\mathbf y}}

\def\v{{\mathbf v}}

\def\cL{\mathcal{L}}

\begin{document}


\title{Assessing coupling dynamics from an ensemble of time series}

\author{Germ{\'a}n G{\'o}mez-Herrero$^\star$}
\author{Wei Wu$^\dag$}%
\author{Kalle Rutanen$^\star$}
\author{Miguel C. Soriano$\S$}
\author{Gordon Pipa$^\dag$}
\author{Raul Vicente$^\dag$}

\affiliation{$^\star$Department of Signal Processing, Tampere
University of Technology, PO 553, 33101 Tampere, Finland}
\affiliation{$^\dag$Max Planck Institute for Brain Research, 60528
Frankfurt am Main, Germany}
\affiliation{$^\S$IFISC, Instituto de F\'{\i}sica Interdisciplinar y
Sistemas Complejos, E-07122, Palma de Mallorca, Spain}

\date{\today}

\begin{abstract}
Finding interdependency relations between (possibly multivariate) time
series provides valuable knowledge about the processes that generate the signals.
Information theory sets a natural framework for non-parametric
measures of several classes of statistical dependencies. However,
a reliable estimation from information-theoretic functionals is hampered
when the dependency to be assessed is brief or evolves in time. Here, we
show that these limitations can be overcome when we have access to an
ensemble of independent repetitions of the time series. In particular,
we gear a data-efficient estimator of probability densities to make
use of the full structure of trial-based measures. By doing so, we can 
obtain time-resolved estimates for a family of \emph{entropy combinations}
(including mutual information, transfer entropy, and their conditional
counterparts) which are more accurate than the simple
average of individual estimates over trials. We show with simulated
and real data that the proposed approach allows to recover the time-resolved
dynamics of the coupling between different subsystems.
\end{abstract}

\pacs{Valid PACS appear here}

\maketitle

A technical problem that arises in various fields of science is to detect interdependencies between
simultaneously measured time series. Namely, the detection of
interdependencies is often the first step for elucidating how the subsystems
that underly the time series interact. For example, in neuroscience,
ecology, or econometrics this approach has lead to the discovery of new
neural codes \cite{gra89.1}, better models of population
dynamics \cite{bjo01.1}, and methods to assess the influence of an
economic variable \cite{gra64.1}, respectively. Tools to unveil an
interdependency include linear techniques, such as cross-correlation
and coherency analysis \cite{gra69.1}, non-linear synchrony measures
\cite{qui02.1}, and the evaluation of statistical dependencies via
mutual information (MI) \cite{cov06.1}. Although useful for assesing the strength of the interaction, these indices do not allow to identify directionality (i.e. cause-response relationships). The latter are arguably more important, if one aims to understand the functioning of a system at the mechanistic level. In this paper we propose a method to reliably estimate the temporal course of directed interactions, as
revealed by several information-theoretic functionals.

While causality is a broad concept, Wiener formulated an operative
definition leaning on the idea that the cause occurs before the effect
and, therefore, knowledge of the cause helps forecasting the effect
\cite{wie56.1}. The widely used Granger causality is the mathematical
formalization of Wiener's definition in terms of linear regressions
\cite{gra69.1}. Alternatively, when a model of the underlying dynamics and of the
interaction is not available, a sound non-parametric approach can be
stated in terms of information theory. A prototypical example is transfer
entropy (TE), which quantifies, in terms of a Kullback-Leibler divergence, how much the
present and past of one system condition (i.e., cause in a Wiener
sense) the dynamics of another \cite{sch00.1}. Nevertheless, the
non-parametric or model-freeness nature of a measure does not usually
come for free. A practical pitfall, common to most
information-theoretic approaches, is that they require many observations to be reliably estimated. This requisite directly confronts
with situations in which the dependency to be analyzed evolves in time
or is subjected to fast transients. When the non-stationarity is only due to a slow
change of a parameter, over-embedding techniques can partially solve
the problem by capturing the slow dynamics of the parameter as an
additional variable \cite{kan04.1}. It is also habitual to de-trend
the time series or divide them into small windows within which the
signals can be considered as approximately stationary. However, the
above-mentioned procedures become unpractical when the relevant
interactions change in a fast time scale. This is the common situation
in brain responses and other complex systems where an external stimuli
elicit a rapid functional reorganization of information-processing
pathways.

Fortunately, in several disciplines the experiments leading to the
multivariate time series can be systematically repeated. Thus, a
typical experimental paradigm might render an ensemble of presumably
independent repetitions or trials
per experimental condition. In this letter we show that this multi-trial
nature can be exploited to produce time-resolved estimates
for a family of information-theoretic measures that we call \emph{entropy combinations}. This family includes well-known functionals such as MI, TE, and their conditional counterparts: partial mutual information~(PMI)~\cite{frenzel07} and partial transfer entropy~(PTE)~\cite{verdes05,gomezherrero10phdthesis}. We use simulations and experimental data to demonstrate that the proposed \emph{ensemble estimators} of entropy combinations are much more accurate than simple averaging of individual trial estimates.


We consider three simultaneously measured time series generated from stochastic processes $X$, $Y$, and $Z$ which can be approximated as stationary Markov processes~\cite{rag02.1} of finite order. The state space of $X$ can then be reconstructed using the delay embedded vectors $\x(n)=\left(x(n),...,x(n-d_x+1)\right)$ for $n=1,\ldots,N$, where $n$ is a time index and $d_x$ is the corresponding Markov order. Similarly we could construct $\y(n)$ and $\z(n)$ for processes $Y$ and $Z$, respectively. Let $V=\left(V_1, ..., V_m\right)$ denote a random $m$-dimensional vector. Then, an \emph{entropy combination} is defined by:

\begin{equation}\label{eq:entropycombination}
C(V_{\cL_1},...,V_{\cL_p}) = \sum_{i=1}^p s_iH(V_{\cL_i})-H(V)
\end{equation}

\noindent where $\forall i \in [1,p]:\;\mathcal{L}_i\subset [1,m]$ and $s_i\in\{-1,1\}$ such that $\sum_{i=1}^{p} s_i \chi_{\cL_i}=\chi_{[1,m]}$ where $\chi_{\mathcal{S}}$ is the characteristic function of a set $\mathcal{S}$. It can be easily checked that MI, TE, PMI and PTE are all entropy combinations:

\begin{eqnarray}
I_{X\leftrightarrow Y}&\equiv & -H_{XY}+H_X+H_Y \nonumber\\
T_{X\leftarrow Y}&\equiv & -H_{WXY}+H_{WX}+H_{XY}-H_{X}\nonumber\\
I_{X\leftrightarrow Y|Z}&\equiv & -H_{XZY}+H_{XZ}+H_{ZY}-H_Z\nonumber\\
T_{X\leftarrow Y|Z}&\equiv & -H_{WXZY}+H_{WXZ}+H_{XZY}-H_{XZ}\nonumber
\end{eqnarray}

\noindent where $W\equiv X^+ \equiv x(n+1)$ so that $H_{WX}$ is the differential entropy of $p(x(n+1),\x(n))$. The latter denotes the joint probability of finding $X$ at states $x(n+1),x(n),...,x(n-d_x+1)$ during time instants $n+1, n, n-1, ..., n-d_x+1$. Notice that, due to stationarity, $p(x(n+1),\x(n))$ is invariant under variations of the time index $n$.

A straightforward approach to the estimation of entropy combinations would be to add separate estimates of each of the involved multi-dimensional entropies. Popular estimators of differential entropy include \emph{plug-in} estimators and fixed and adaptive histogram or partition methods. However, other non-parametric techniques such as kernel and nearest-neighbor estimators have been shown to be extremely more data-efficient and accurate~\cite{vic02.1}. An asymptotically unbiased estimator based on nearest-neighbor statistics is by Kozachenko and Leonenko (KL)~\cite{koz87.1}. For $N$ realizations $\x[1],\x[2],...,\x[N]$ of a $d$-dimensional random vector $X$, the KL estimator takes the form:

\begin{equation}\label{eq:klestimator}
\hat{H}_X = -\psi(k)+\psi(N)+\log(v_d)+\frac{d}{N}\sum_{i=1}^N\log (\epsilon(i))
\end{equation}

\noindent \noindent where $\psi$ is the digamma function, $v_{d}$ is the volume of the $d$-dimensional unit ball, and $\epsilon(i)$ is the distance from $\x[i]$ to its kth nearest neighbor in the set $\left\{\x[j]\right\}_{\forall j\neq i}$. The KL estimator is based on the assumption that the density of the distribution of random vectors is constant within an $\epsilon$-ball. The bias of the final entropy estimate depends on the validity of this assumption, and thus, on the values of $\epsilon(n)$. Since the size of the $\epsilon$-balls depends directly on the dimensionality of the random vector, the biases of estimates for the differential entropies in~(\ref{eq:entropycombination}) will, in general, not cancel, leading to a poor estimator of the entropy combination. This problem can be partially overcome by noticing that~(\ref{eq:klestimator}) holds for any value of $k$ so that we do not need to have a fixed $k$. Therefore, we can vary the value of $k$ in each data point so that the radius of the corresponding $\epsilon$-balls would be approximately the same for the joint and the marginal spaces. This idea was originally proposed in~\cite{kra04.1} for estimating mutual information, was used in~\cite{fre07.1} to estimate PMI, and we generalize it here to the following estimator of \emph{entropy combinations}:

\begin{equation}\label{eq:entropycombinationestimator}
\hat{C}(V_{\cL_1},...,V_{\cL_p}) = F(k)-\sum_{i=1}^{p}s_i\left<F\left(k_i(n)\right)\right>_{n}
\end{equation}

\noindent where $F(k)=\psi(k)-\psi(N)$ and $\left<\cdots \right>_n=\frac{1}{N}\sum_{n=1}^{N}(\cdots)$ denotes averaging with respect to the time index. The term $k_i(n)$ accounts for the number of neighbors of the $n$th realization of the marginal vector $V_{\mathcal{L}_i}$ located at a distance strictly less than $\epsilon(n)$, where $\epsilon(n)$ denotes the radius of the $\epsilon$-ball in the joint space. The point itself is included in this counting. 

A fundamental limitation of estimator~(\ref{eq:entropycombinationestimator}) is the assumption that the involved multidimensional distributions are stationary. However, this is hardly the case in many real applications and time-adaptation becomes crucial in order to obtain meaningful estimates. A trivial solution is to use the following time-varying estimator of entropy combinations:

\begin{equation}
\hat{C}(\{V_{\cL_1},...,V_{\cL_p}\},n) = F(k)-\sum_{i=1}^{p}s_i F\left(k_i(n)\right)
\end{equation} 

\begin{figure}
\includegraphics[height=.20\textheight]{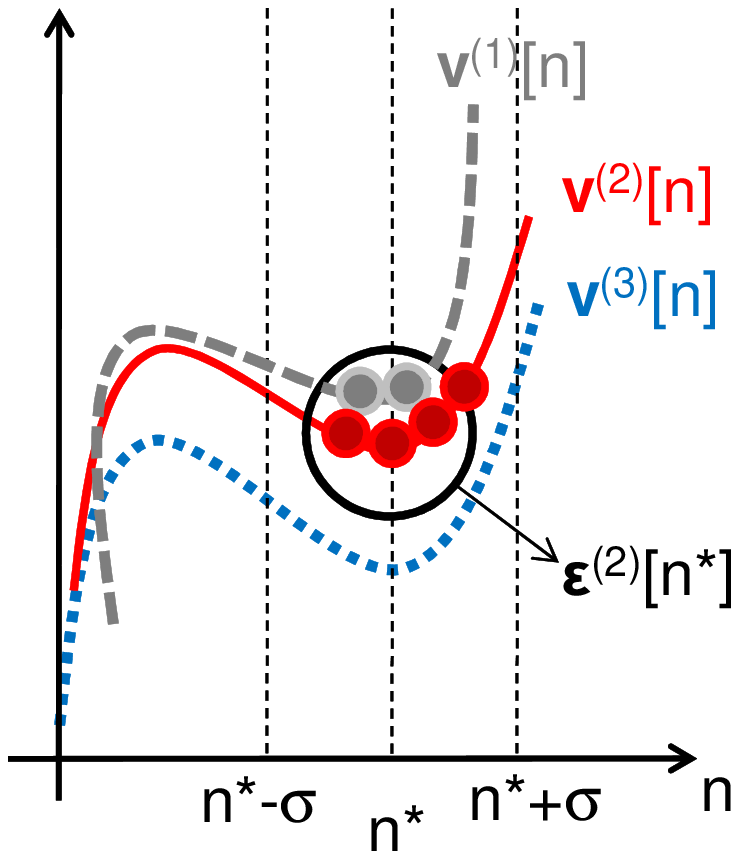}
\includegraphics[height=.20\textheight]{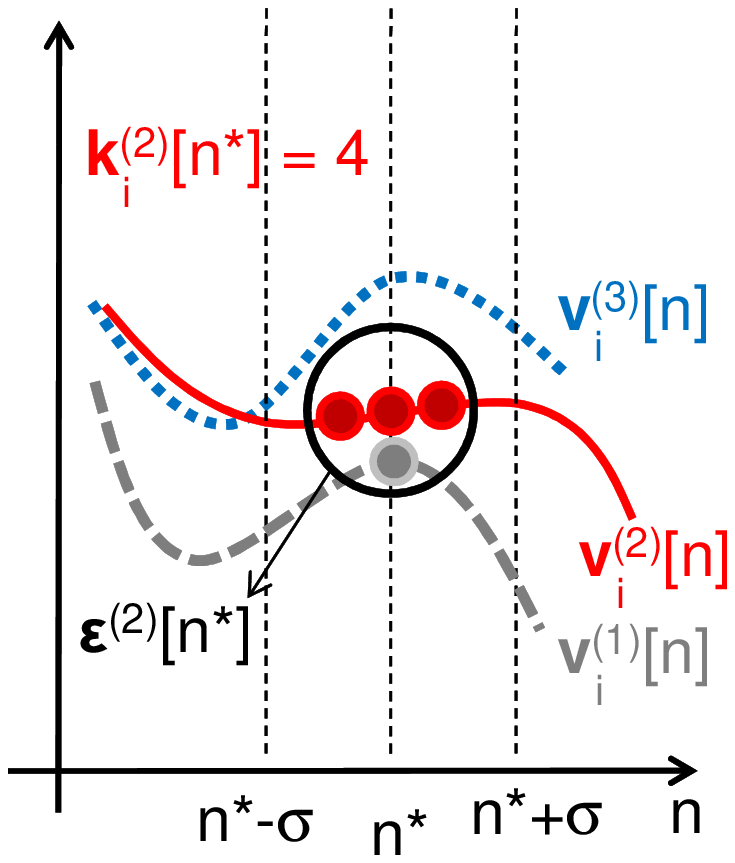}
\caption{Nearest neighbor statistics across trials. a) For each time instant $n=n^*$ and trial $r=r^*$, we compute the (maximum norm) distance $\epsilon^{(r^*)}(n^*)$ from $\v^{(r^{*})}[n^*]$ to its $k$-th nearest neighbor among all trials. Here the procedure is illustrated for $k=5$. b) $k_i^{(r^*)}[n^*]$ counts how many neighbors of $\v_i^{(r^*)}[n^*]$ are within a radius $\epsilon^{r^*}(n^*)$. The point itself (i.e. $\v_i^{(r^*)}[n^*]$) is also included in this count. These neighbor counts are obtained for all $i=1,...p$ marginal trajectories.}
\label{fig:ensembleneighbors}
\end{figure}

This naive time-adaptive estimator is not useful in practice due to its large variance, which stems from the fact that a single data point is used for producing the estimate at each time instant. However, let us consider the case of an ensemble of $r'$ repeated measurements (trials) from the dynamics of $V$. Let us also denote by $\left\{\v^{(r)}[n]\right\}_r$ the measured dynamics for those trials ($r=1,2,...r'$). Similarly, we denote by $\{\v_i^{(r)}[n]\}_r$ the measured dynamics for the marginal vector $V_{\mathcal{L}_i}$. A straightforward approach for integrating the information from different trials is to average together estimates obtained from individual trials:

\begin{equation}\label{eq:avgestimator}
\hat{C}^{\textrm{avg}}(\{V_{\cL_1},...,V_{\cL_p}\},n)= \frac{1}{r'}\sum_{r=1}^{r'}\hat{C}^{(r)}(\{V_{\cL_1},...,V_{\cL_p}\},n)\nonumber
\end{equation}
where $\hat{C}^{(r)}(\{V_{\cL_1},...,V_{\cL_p}\},n)$ is the estimate obtained from the $r$th trial. However, this approach makes a poor use of the available data and will typically produce useless estimates, as will be shown in the experimental section of this chapter. A more effective procedure takes into account the multi-trial nature of our data by searching for neighbors across ensemble members, rather than from within each individual trial. This \emph{nearest ensemble neighbors}~\cite{kramer04} approach is illustrated in Fig.~\ref{fig:ensembleneighbors} and leads to the following ensemble estimator of entropy combinations:

\begin{equation}\label{eq:ensembleestimator}
\hat{C}^{\textrm{en}}(\{V_{\cL_1},...,V_{\cL_p}\},n)= F(k)-\frac{1}{r'}\sum_{r=1}^{r'}\sum_{i=1}^{p}s_i F\left(k_i^{(r)}(n)\right)\nonumber
\end{equation}

\noindent where the counts of marginal neighbors $\{k_i^{(r)}(n)\}_{\forall i=1,...p}^{\forall  r=1,...,r'}$ are computed using overlapping time-windows of size $2\sigma$, as shown in Fig.~\ref{fig:ensembleneighbors}. For rapidly changing connectivity patterns, small values of $\sigma$ might be needed to track the coupling dynamics while larger values of $\sigma$ will lead to lower estimator variance.

To demonstrate that $\hat{\Theta}^{\textrm{en}}$ can be used to characterize dynamic coupling patterns we simulated three non-linearly coupled autoregressive processes with a time-varying coupling factor:

\[
\begin{array}{lll}
x^r[n] &=& 0.4 x^r[n-1]+\eta_x \, ,\\
y^r[n] &=& 0.5 y^r[n-1]+\kappa_{yx}[n]\sin\left(x^r[n-\tau_{yx}]\right)+\eta_y \, ,\\
z^r[n] &=& 0.5 z^r[n-1]+\kappa_{zy}[n]\sin\left(y^r[n-\tau_{zy}]\right)+\eta_z \, .
\end{array}
\]

\noindent during 1500 time steps and repeated R=50 trials with new initial conditions. The terms $\eta_x$, $\eta_y$ and $\eta_z$ represent normally distributed noise processes, which are mutually independent across trials and time instants. The coupling delays amount to $\tau_{yx}=10$, $\tau_{zy}=15$ while the dynamics of the coupling follows a sinusoidal variation:

\[
k_{yx}[n] = \left\{
\begin{array}{ll}
\sin\left(\frac{2\pi n}{500}\right) & \textrm{for} \; 250 \leq n < 750\\
0 & \textrm{otherwise}\\
\end{array}
\right.
\]

\[
k_{zy}[n] = \left\{
\begin{array}{ll}
\cos\left(\frac{2\pi n}{500}\right) & \textrm{for} \; 750 \leq n < 1250\\
0 & \textrm{otherwise}\\
\end{array}
\right.
\]

\begin{figure}
\includegraphics[width=.47\textwidth]{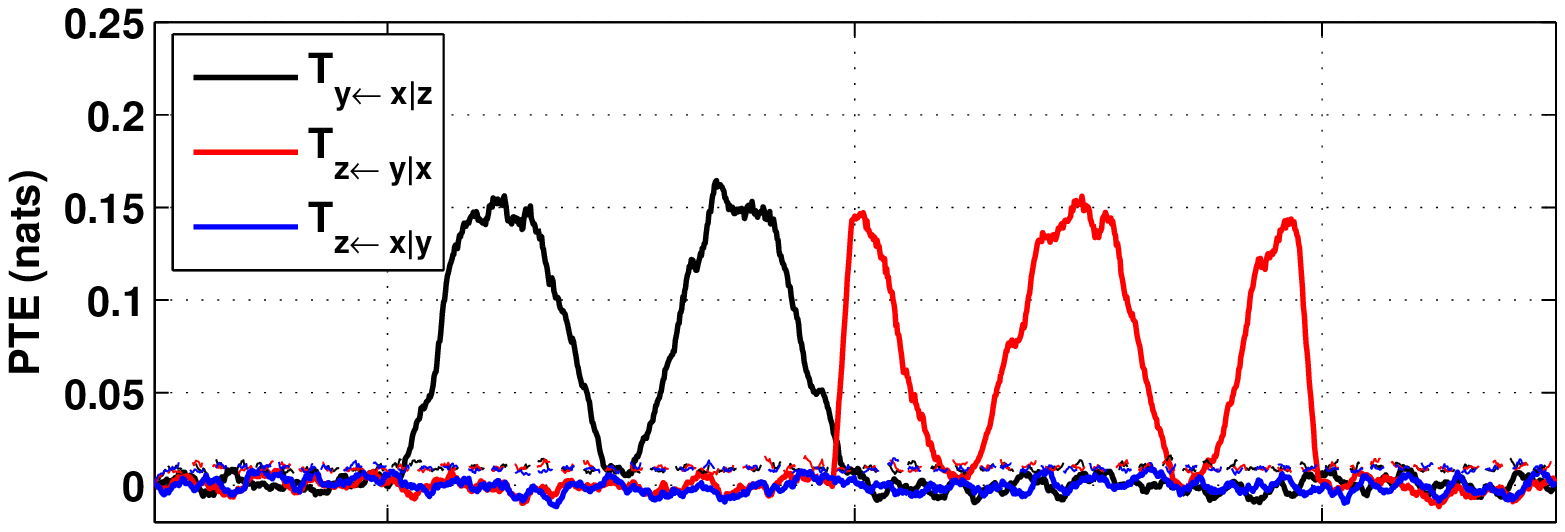}
\includegraphics[width=.47\textwidth]{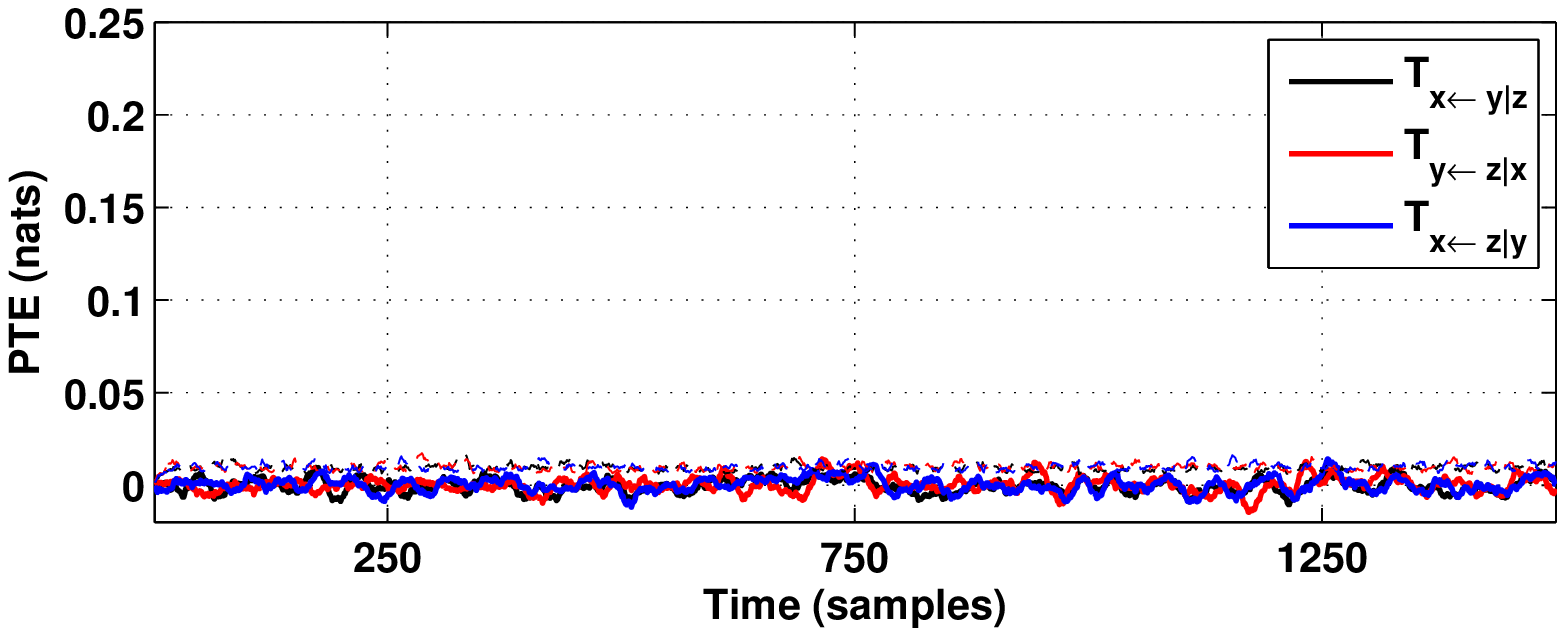}
\caption{Partial transfer entropy between three non-linearly coupled Gaussian processes. The solid lines represent PTE values while the color-matched dashed lines denote corresponding $p=0.05$ significance levels. The temporal variance of the PTE estimates was reduced with a post-processing moving average filter of order 20.}
\label{fig:tvcorrgauss}
\end{figure}

Before PTE estimation each time-series was time-delayed so that they had maximal mutual information with the destination of the flow. That is, before computing some $T_{a\leftarrow b|c}(n)$, the time-series $b$ and $c$ were delayed so that they shared maximum information with the time-series $a$. To assess the statistical significance of the PTE values (at each time-instant) we applied a permutation test with surrogate data generated by randomly shuffling trials~\cite{pes01.1}. Fig.~\ref{fig:tvcorrgauss} shows the time-varying PTEs obtained for these data with the ensemble estimator of entropy combinations given in Eq.~\ref{eq:ensembleestimator}. Indeed, the PTE analysis accurately describes the underlying interaction dynamics. In particular, it captures both the onset/offset and the oscillatory profile of the effective coupling across the three processes. On the other hand, the naive average estimator~(\ref{eq:avgestimator}) did not reveal any significant flow of information between the three time-series (see supplementary material).

\begin{figure}[t]
\centering
\includegraphics[width=0.48\textwidth]{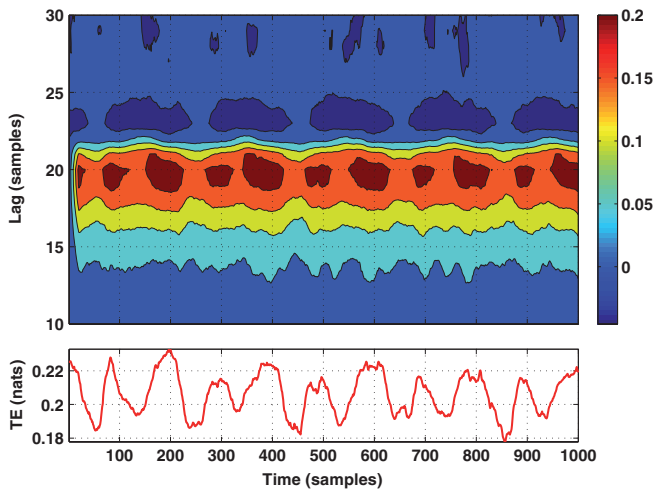}
\caption{Transfer entropy from the first electronic circuit towards the second. The upper figure shows time-varying TE versus the lag introduced in the temporal activation of the first circuit. Clearly, there is a directional flow of information time-locked at lag $\tau=20$ samples, which is significant for all time-instants ($p<0.01$). On the other hand, the flow of information in the opposite direction was much smaller ($T_{1\leftarrow 2}<0.0795$ nats $\forall (t,\tau)$) and only reached significance ($p<0.01$) for about $1\%$ of the tuplas $(n,\tau)$. The lower figure shows that the temporal pattern of information flow for $\tau=20$, i.e. $T_{2\leftarrow 1}(n,\tau=20)$, which resembles a sinusoid with a period of roughly 100 data samples.}
\label{fig:realexample}
\end{figure}

To evaluate the robustness and performance of the entropy combination estimator to real levels of noise and measurements variability, we also present a second example derived from experimental data on electronic circuits. The system consists of two nonlinear Mackey-Glass circuits unidirectionally coupled through their voltage variables. The master circuit is additionally subject to a feedback loop responsible for generating high dimensional chaotic dynamics. A time-varying effective coupling is then induced by periodically modulating the strength of the coupling between circuits as controlled by an external CPU. In this case, we applied transfer entropy between the voltage signals generated from the two circuits for 180 trials, each 1000 sampling times long. Figure ~\ref{fig:realexample} shows the TE estimates obtained with~(\ref{eq:ensembleestimator}) versus the temporal lag introduced between the two voltage signals (intended to scan the unknown coupling delay). The results show that the TE estimates capture perfectly the dynamics of the effect exerted by the master circuit on the slave circuit. On the other hand, no significant coupling is detected in the reverse direction (see supplementary material). Both the period of the coupling dynamics (100 samples) and the coupling delay (20 samples) can be accurately recovered from Fig.~\ref{fig:realexample}.

In conclusion, we have introduced an ensemble estimator of \emph{entropy combinations} that is able to detect time-varying information flow between dynamical systems, provided that an ensemble of repeated measurements is available for each system. The proposed approach allows to construct time-adaptive estimators of MI, PMI, TE and PTE, which are the most common information-theoretic measures for dynamical coupling analyses. Using simulations and real physical measurements from electronic circuits we showed that these new estimators can accurately describe multivariate coupling dynamics. It is important to mention that intrinsic to our approach is the assumption that the evolution of the interdependencies to be detected are to some degree "locked" to the trial onset. This is typically the case when some controlled external perturbation induce or evokes the interactions across the subsystems measured. The degree of locking determines the maximum temporal resolution achievable by the method (which is controlled via $\sigma$). Nevertheless, alignment techniques can help to reduce the possible jitter across trials and thus increase the resolution. The methods presented here are general but we anticipate that a potential application is the analysis of the mechanisms underlying the generation of event-related brain responses and the seasonal variations of geophysical variables. To promote dissemination we have publicly released a software library that includes efficient implementations of these and other information-theoretic methods~\cite{timtoolbox}.

This work has been supported by the EU project GABA (FP6-2005-NEST-Path 043309) and by the Finnish Foundation for Technology Promotion.


\begin{thebibliography}{20}
\expandafter\ifx\csname natexlab\endcsname\relax\def\natexlab#1{#1}\fi
\expandafter\ifx\csname bibnamefont\endcsname\relax
  \def\bibnamefont#1{#1}\fi
\expandafter\ifx\csname bibfnamefont\endcsname\relax
  \def\bibfnamefont#1{#1}\fi
\expandafter\ifx\csname citenamefont\endcsname\relax
  \def\citenamefont#1{#1}\fi
\expandafter\ifx\csname url\endcsname\relax
  \def\url#1{\texttt{#1}}\fi
\expandafter\ifx\csname urlprefix\endcsname\relax\def\urlprefix{URL }\fi
\providecommand{\bibinfo}[2]{#2}
\providecommand{\eprint}[2][]{\url{#2}}

\bibitem[{\citenamefont{Gray et~al.}(1989)\citenamefont{Gray, Konig, Engel, and
  Singer}}]{gra89.1}
\bibinfo{author}{\bibfnamefont{C.}~\bibnamefont{Gray}},
  \bibinfo{author}{\bibfnamefont{P.}~\bibnamefont{Konig}},
  \bibinfo{author}{\bibfnamefont{A.}~\bibnamefont{Engel}}, \bibnamefont{and}
  \bibinfo{author}{\bibfnamefont{W.}~\bibnamefont{Singer}},
  \bibinfo{journal}{Nature} \textbf{\bibinfo{volume}{338}},
  \bibinfo{pages}{334} (\bibinfo{year}{1989}).

\bibitem[{\citenamefont{Bjornstad and Grenfell}(2001)}]{bjo01.1}
\bibinfo{author}{\bibfnamefont{O.}~\bibnamefont{Bjornstad}} \bibnamefont{and}
  \bibinfo{author}{\bibfnamefont{B.}~\bibnamefont{Grenfell}},
  \bibinfo{journal}{Science} \textbf{\bibinfo{volume}{293}},
  \bibinfo{pages}{638} (\bibinfo{year}{2001}).

\bibitem[{\citenamefont{Granger and Hatanaka}(1964)}]{gra64.1}
\bibinfo{author}{\bibfnamefont{C.}~\bibnamefont{Granger}} \bibnamefont{and}
  \bibinfo{author}{\bibfnamefont{M.}~\bibnamefont{Hatanaka}},
  \emph{\bibinfo{title}{Spectral analysis of economic time series}}
  (\bibinfo{publisher}{Princeton University Press}, \bibinfo{year}{1964}).

\bibitem[{\citenamefont{Granger}(1969)}]{gra69.1}
\bibinfo{author}{\bibfnamefont{C.}~\bibnamefont{Granger}},
  \bibinfo{journal}{Econometrica} \textbf{\bibinfo{volume}{37}},
  \bibinfo{pages}{424} (\bibinfo{year}{1969}).

\bibitem[{\citenamefont{Quian~Quiroga et~al.}(2002)\citenamefont{Quian~Quiroga,
  Kraskov, Kreuz, and Grassberger}}]{qui02.1}
\bibinfo{author}{\bibfnamefont{R.}~\bibnamefont{Quian~Quiroga}},
  \bibinfo{author}{\bibfnamefont{A.}~\bibnamefont{Kraskov}},
  \bibinfo{author}{\bibfnamefont{T.}~\bibnamefont{Kreuz}}, \bibnamefont{and}
  \bibinfo{author}{\bibfnamefont{P.}~\bibnamefont{Grassberger}},
  \bibinfo{journal}{Phys. Rev. E} \textbf{\bibinfo{volume}{65}},
  \bibinfo{pages}{041903} (\bibinfo{year}{2002}).

\bibitem[{\citenamefont{Cover and Thomas}(2006)}]{cov06.1}
\bibinfo{author}{\bibfnamefont{T.}~\bibnamefont{Cover}} \bibnamefont{and}
  \bibinfo{author}{\bibfnamefont{J.}~\bibnamefont{Thomas}},
  \emph{\bibinfo{title}{Elements of information theory}}
  (\bibinfo{publisher}{Wiley}, \bibinfo{year}{2006}).

\bibitem[{\citenamefont{Wiener}(1956)}]{wie56.1}
\bibinfo{author}{\bibfnamefont{N.}~\bibnamefont{Wiener}},
  \emph{\bibinfo{title}{Modern mathematics for engineers}}
  (\bibinfo{publisher}{McGraw-Hill, New York}, \bibinfo{year}{1956}), chap.
  \bibinfo{chapter}{The theory of prediction}.

\bibitem[{\citenamefont{Schreiber}(2000)}]{sch00.1}
\bibinfo{author}{\bibfnamefont{T.}~\bibnamefont{Schreiber}},
  \bibinfo{journal}{Phys. Rev. Lett.} \textbf{\bibinfo{volume}{85}},
  \bibinfo{pages}{461} (\bibinfo{year}{2000}).

\bibitem[{\citenamefont{Kantz and Schreiber}(2004)}]{kan04.1}
\bibinfo{author}{\bibfnamefont{H.}~\bibnamefont{Kantz}} \bibnamefont{and}
  \bibinfo{author}{\bibfnamefont{T.}~\bibnamefont{Schreiber}},
  \emph{\bibinfo{title}{Nonlinear time series analysis}}
  (\bibinfo{publisher}{Cambridge university press}, \bibinfo{year}{2004}),
  \bibinfo{edition}{2nd} ed.

\bibitem[{\citenamefont{Frenzel and Pompe}(2007{\natexlab{a}})}]{frenzel07}
\bibinfo{author}{\bibfnamefont{S.}~\bibnamefont{Frenzel}} \bibnamefont{and}
  \bibinfo{author}{\bibfnamefont{B.}~\bibnamefont{Pompe}},
  \bibinfo{journal}{Phys. Rev. Lett.} \textbf{\bibinfo{volume}{99}},
  \bibinfo{pages}{204101} (\bibinfo{year}{2007}{\natexlab{a}}).

\bibitem[{\citenamefont{Verdes}(2005)}]{verdes05}
\bibinfo{author}{\bibfnamefont{P.~F.} \bibnamefont{Verdes}},
  \bibinfo{journal}{Phys. Rev. E} \textbf{\bibinfo{volume}{72}},
  \bibinfo{pages}{026222} (\bibinfo{year}{2005}).

\bibitem[{\citenamefont{G{\'o}mez-Herrero}(2010)}]{gomezherrero10phdthesis}
\bibinfo{author}{\bibfnamefont{G.}~\bibnamefont{G{\'o}mez-Herrero}}, Ph.D.
  thesis, \bibinfo{school}{Tampere University of Technology, Department of
  Signal Processing} (\bibinfo{year}{2010}).

\bibitem[{\citenamefont{Ragwitz and Kantz}(2002)}]{rag02.1}
\bibinfo{author}{\bibfnamefont{M.}~\bibnamefont{Ragwitz}} \bibnamefont{and}
  \bibinfo{author}{\bibfnamefont{H.}~\bibnamefont{Kantz}},
  \bibinfo{journal}{Physical Review E} \textbf{\bibinfo{volume}{65}},
  \bibinfo{pages}{056201} (\bibinfo{year}{2002}).

\bibitem[{\citenamefont{Victor}(2002)}]{vic02.1}
\bibinfo{author}{\bibfnamefont{J.~D.} \bibnamefont{Victor}},
  \bibinfo{journal}{Phys. Rev. E} \textbf{\bibinfo{volume}{66}},
  \bibinfo{pages}{051903} (\bibinfo{year}{2002}).

\bibitem[{\citenamefont{Kozachenko and Leonenko}(1987)}]{koz87.1}
\bibinfo{author}{\bibfnamefont{L.}~\bibnamefont{Kozachenko}} \bibnamefont{and}
  \bibinfo{author}{\bibfnamefont{N.}~\bibnamefont{Leonenko}},
  \bibinfo{journal}{Problemy Peredachi Informatsii}
  \textbf{\bibinfo{volume}{23}}, \bibinfo{pages}{9} (\bibinfo{year}{1987}).

\bibitem[{\citenamefont{Kraskov et~al.}(2004)\citenamefont{Kraskov,
  St\"ogbauer, and Grassberger}}]{kra04.1}
\bibinfo{author}{\bibfnamefont{A.}~\bibnamefont{Kraskov}},
  \bibinfo{author}{\bibfnamefont{H.}~\bibnamefont{St\"ogbauer}},
  \bibnamefont{and}
  \bibinfo{author}{\bibfnamefont{P.}~\bibnamefont{Grassberger}},
  \bibinfo{journal}{Phys. Rev. E} \textbf{\bibinfo{volume}{69}},
  \bibinfo{pages}{066138} (\bibinfo{year}{2004}).

\bibitem[{\citenamefont{Frenzel and Pompe}(2007{\natexlab{b}})}]{fre07.1}
\bibinfo{author}{\bibfnamefont{S.}~\bibnamefont{Frenzel}} \bibnamefont{and}
  \bibinfo{author}{\bibfnamefont{B.}~\bibnamefont{Pompe}},
  \bibinfo{journal}{Phys. Rev. Lett.} \textbf{\bibinfo{volume}{99}},
  \bibinfo{pages}{204101} (\bibinfo{year}{2007}{\natexlab{b}}).

\bibitem[{\citenamefont{Kramer et~al.}(2004)\citenamefont{Kramer, Edwards,
  Soltani, Berger, Knight, and Szeri}}]{kramer04}
\bibinfo{author}{\bibfnamefont{M.~A.} \bibnamefont{Kramer}},
  \bibinfo{author}{\bibfnamefont{E.}~\bibnamefont{Edwards}},
  \bibinfo{author}{\bibfnamefont{M.}~\bibnamefont{Soltani}},
  \bibinfo{author}{\bibfnamefont{M.~S.} \bibnamefont{Berger}},
  \bibinfo{author}{\bibfnamefont{R.~T.} \bibnamefont{Knight}},
  \bibnamefont{and} \bibinfo{author}{\bibfnamefont{A.~J.} \bibnamefont{Szeri}},
  \bibinfo{journal}{Phys. Rev. E.} \textbf{\bibinfo{volume}{70}},
  \bibinfo{pages}{011914} (\bibinfo{year}{2004}).

\bibitem[{\citenamefont{Pesarin}(2001)}]{pes01.1}
\bibinfo{author}{\bibfnamefont{F.}~\bibnamefont{Pesarin}},
  \emph{\bibinfo{title}{Multivariate permutation tests}}
  (\bibinfo{publisher}{John Wiley and Sons}, \bibinfo{year}{2001}).

\bibitem[{\citenamefont{Rutanen}()}]{timtoolbox}
\bibinfo{author}{\bibfnamefont{K.}~\bibnamefont{Rutanen}},
  \emph{\bibinfo{title}{{TIM} {C++} library}}, \bibinfo{howpublished}{Available
  online: http://www.tut.fi/tim}.

\end{thebibliography}
\end{document}